\documentclass[prl,letterpaper,twocolumn,showpacs,superscriptaddress,floatfix]{revtex4}
\usepackage{graphicx,psfrag,amsmath,amssymb,amsfonts,bbm,latexsym,color,dcolumn,bm}

\begin{document}

\title{Sympathetic cooling route to Bose-Einstein condensate and Fermi-liquid mixtures}

\author{Robin C\^ot\'e}
\affiliation{Department of Physics, University of Connecticut, Storrs,
CT 06269, USA}

\author{Roberto Onofrio}
\affiliation
{\mbox{Department of Physics and Astronomy,
Dartmouth College, 6127 Wilder Laboratory, Hanover, NH 03755, USA}}
\affiliation{Dipartimento di Fisica ``G. Galilei'',
    Universit\`a di Padova, Via Marzolo 8, Padova 35131, Italy}
\affiliation{Center for Statistical Mechanics and Complexity, INFM,
Unit\`a di Roma 1, Roma 00185, Italy}

\author{Eddy Timmermans}
\affiliation{T-4, Theory Division, Los Alamos National Laboratory,
Los Alamos, NM 87545, USA}

\begin{abstract}
We discuss a sympathetic cooling strategy that can successfully
mitigate fermion-hole heating in a dilute atomic Fermi-Bose mixture
and access the temperature regime in which the fermions behave as a
Fermi-liquid.  We introduce an energy-based formalism to describe
the temperature dynamics with which we study a specific and promising
mixture composed of $^{6}$Li and $^{87}$Rb.  Analyzing the harmonically
trapped mixture, we find that the favourable features of this mixture
are further enhanced by using different trapping frequencies for the 
two species.
\end{abstract}
\maketitle

\noindent
{\it Introduction.}  Cold atom mixtures of fermions and Bose-Einstein 
condensates (BEC) \cite{sympat} suggest a new venue for the study of 
quantum liquid mixtures that were previously available only in the 
form of condensed $^{3}$He-$^{4}$He mixtures \cite{Baym}. 
While we understand the helium phase separation to be caused by
mediated interactions \cite{BBP66}, a first-principle description 
of this and other helium quantities is complicated by strong 
interaction effects \cite{Krotschek}.  
These also reduce the boson mediated interactions that can Cooper-pair 
\cite{BBP66} the $^{3}$He fermions, giving a critical temperature 
$T_{c}$ currently unaccessible. Mediated interactions and their 
prominence near a quantum phase transition (quantum criticality, 
see \cite{Stewart}) have also been identified as the probable 
cause of the observed non-Fermi-liquid behavior in high 
$T_{c}$ superconductors \cite{Stewart} and, perhaps, of 
high $T_{c}$ superconductivity itself.  
In these systems, however, a clear understanding is again 
impeded by the ambiguity of experimental data.

The above issues \cite{lump} can be studied in a much cleaner and 
more accessible environment when cold atom physicists succeed in cooling 
fermion-boson mixtures into the Fermi-liquid regime \cite{Baym} in 
which the fermions behave in a universal manner that
follows from expanding the energy to second order in quasi-particle 
occupation numbers.  This implies a heat capacity that varies 
linearly with temperature, a feature that often serves as
experimental test of Fermi-liquid behavior.  
In a dilute fermion gas, it sets in below ten percent of the 
Fermi temperature $T<0.1 T_{F}$, whereas $T<0.02 T_{F}$ is required in
the $^{3}$He liquid.  For cold atom quantum liquid studies, which can 
also access fermion-boson mixtures in a parameter regime that is 
fundamentally different from that of the helium mixtures \cite{us}, 
crossing into the Fermi-liquid regime is as crucial of a step as
cooling below $T_{c}$ is for the study of superfluidity.  So far, 
experiments that use bosons to cool fermion atoms (sympathetic
cooling) have not entered this regime \cite{EXPT0102} because the 
low temperature drop in the BEC heat capacity impairs its ability 
to refrigerate \cite{Onofrio}, and because the unavoidable
loss of fermions leads to significant heating \cite{Eddy2}.  
The BEC introduces further losses in the Fermi system (three-body 
recombinations) thereby increasing the Fermi-hole heating rate.

In this Rapid, we show that with a large boson-to-fermion mass ratio
and with the proper trapping parameters, sympathetic cooling {\it can}
access the Fermi-liquid temperature regime.  Heavy bosons require
many collisions from the lighter fermions before they heat significantly.
The corresponding increase in BEC heat capacity is crucial; low three-body
recombination rates are helpful.  The need to strike a delicate
balance between keeping the fermion boson overlap sufficiently low
to reduce hole-heating, but sufficiently high to ensure efficient
fermion-BEC heat exchange can be met by adjusting the fermion
and boson trapping frequencies in a bichromatic trapping scheme.
We illustrate these points for the promising candidate of a
$^{6}$Li--$^{87}$Rb mixture, which has been recently experimentally 
investigated \cite{Silber}.

\noindent
{\it Loss-induced Temperature Dynamics.} - The cold atom many-body states
of interest are metastable (the true ground state being a condensed
solid) and trapped atom gases have a finite lifetime. 
Atoms leave the trap as a consequence of collisions with room
temperature background atoms, of two-body spin relaxation processes 
(in magnetic traps), and of three-body recombination.
Instead of describing the many-body response to particle loss,
we keep track of the energy-balance and define the effective
temperature of the system $T$ as the temperature at which an 
equilibrium system with the same number of particles has the same energy.

The above collisions reduce the number of particles of type $j$, 
$N_{j}$, as well as internal energy $E$, as each loss removes the energy
of the lost particles.  If the collision products, on their
way out of the trap, collide with the remaining atoms,
the process also adds an average kinetic energy $\epsilon_{j}^{coll}$.
We equate the total energy balance rate, $d E/dt -\sum_{j}
\epsilon_{j}^{coll} d N_{j}/dt$ to the variation of the total
equilibrium energy of a temperature-relaxed system,
$d E(N_{j},T)/dt = (\partial E / \partial T)
dT/dt + \sum_{j} (\partial E/\partial N_{j}) d N_{j}/dt$ and
we identify the usual thermodynamic derivatives,
$(\partial E / \partial T) = C$, where $C$ denotes
the heat capacity at constant number of particles,
and $(\partial E/\partial N_{j}) = \mu_{j}$, where
$\mu_{j}$ represents the chemical potential of
the $j$-particles.  Equating the energy rates and solving
for the temperature derivative, we obtain the central
equation of temperature dynamics
\begin{equation}
       \frac{d T}{dt} = \frac{1}{C} \left(\frac{dF}{dt} -
       \sum_{j} \epsilon_{j}^{coll} \frac{d N_{j}}{dt}\right) \; ,
\label{tempdyn}
\end{equation}
where we have defined $F$ as the free energy that occurs in 
Fermi-liquid theory: $F = E -\sum_{j} \mu_{j} N_{j}$ 
(here dependent on $N_{j}$ rather than $\mu_{j}$).

\noindent
{\it Homogeneous ideal gas Fermi-Bose mixtures.} -
The homogeneous mixture of a single component degenerate fermion
gas and a BEC allows a transparent description of hole heating and
its mitigation by sympathetic cooling.
Apart from border effects, the homogeneous mixture can be realized
by using blue-detuned laser light sheets \cite{Davidson}.
\begin{figure}[t]
\psfrag{x}[][]{$t/\tau$}
\psfrag{y}[][]{$T/T_\mathrm{F}$}
\psfrag{LiNa}[][]{$^6$Li-$^{23}$Na}
\psfrag{LiRb}[][]{$^6$Li-$^{87}$Rb}
\psfrag{KNa}[][]{$^{40}$K-$^{23}$Na}
\psfrag{KRb}[][]{$^{40}$K-$^{87}$Rb}
\psfrag{Fermi}[][]{Fermi gas}
\includegraphics[width=0.90\columnwidth,clip]{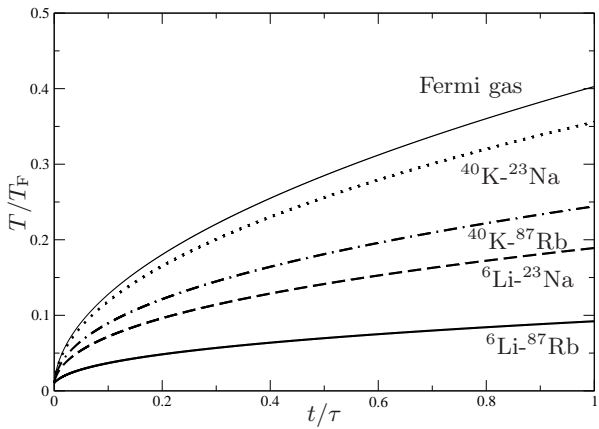}
\caption{Hole heating in homogeneous Fermi-Bose mixtures.
The temperature, $T$, scaled by the Fermi temperature, $T_{F}$, is 
depicted for four Fermi-Bose mixtures consisting of fermionic $^6$Li
or $^{40}$K atoms with $^{23}$Na or $^{87}$Rb as bosonic partners,
starting at $T/T_\mathrm{F}=10^{-2}$.
For comparison, the hole heating curve for pure fermions, assuming
the same loss-rate, is also plotted (lighter continuous curve).}
\label{fig1}
\end{figure}
Atoms in the mixture undergo various three-body loss processes. 
For instance, a fermion can recombine with a boson into a highly 
energetic molecule while ejecting a second boson, corresponding to
\begin{equation}
\dot{n}_\mathrm{B} = -2 \alpha n_\mathrm{B}^2 n_\mathrm{F},\ \ \
\dot{n}_\mathrm{F} =  - \alpha n_\mathrm{B}^2 n_\mathrm{F}
\label{3b}
\end{equation}
where $n_\mathrm{F}$ and $n_\mathrm{B}$ represent the fermion
and boson densities respectively, and $\alpha$ denotes the
three-body loss rate coefficient.  Processes involving three-boson
collisions do not lead to significant heating and those involving 
two or three indistinguishable fermions are Pauli-inhibited.
If, as in Eq. (\ref{3b}), the rate is energy-insensitive, the loss 
of one out of $N_{F}$ fermions removes, on average, a kinetic 
energy $\frac{3}{5} E_\mathrm{{\rm F}}$, where $E_\mathrm{{\rm F}}$
is the Fermi energy.  At the same temperature the remaining 
degenerate system of $N_{{\rm F}}-1$ fermions would have an 
energy $E_\mathrm{{\rm F}}$ less than the initial Fermi system 
of $N_{F}$ fermions.
The net result is an effective energy increase of $\frac{2}{5} 
E_\mathrm{{\rm F}}$ per fermion lost (assuming, optimistically, 
$\epsilon^{{\rm coll}}_{j}=0$). We apply Eq. (\ref{tempdyn}) to 
the homogeneous mixture of fermions of mass $m_{F}$, and an ideal
BEC of $N_{{\rm B}}$ bosons of mass $m_{{\rm B}}$.  
The total heat capacity of an an ideal or near-deal gas mixture 
is the sum of the BEC and fermion heat capacities, $C=k_{B} N_{{\rm F}}
(\pi^{2}/2) (T/T_{{\rm F}}) [1 + (45/8\pi^{3/2}) \zeta_{5/2}(1) 
(m_{{\rm B}}/m_{{\rm F}})^{3/2} \sqrt{T/T_{{\rm F}}}]$, where 
$k_{B}$ denotes the Boltzmann constant, $T_{{\rm F}}$ is the 
Fermi-temperature, $T_{{\rm F}}=E_{{\rm F}}/k_{B}$ and $\zeta_{5/2}(1) = 1.342$.
The time scale of the dynamics is the lifetime $\tau$ of the fermion system,
$\tau^{-1} = \alpha n_{{\rm B}}^{2}$,
\begin{equation}
\frac{d(T/T_\mathrm{F})}{d(t/\tau)}=\frac{\frac{4}{5\pi^2}
\left(\frac{T}{T_\mathrm{F}}\right)^{-1}}
{1+ \frac{45 \zeta_{5/2}(1)}{8\pi^{3/2}}\left(
\frac{m_\mathrm{B}}{m_\mathrm{F}}\right)^{3/2}
\left( \frac{T}{T_\mathrm{F}} \right)^{1/2}}  \; .
   \label{heatinghom}
\end{equation}
The term $\propto \sqrt{T/T_{{\rm F}}}$ in the denominator stems from 
the BEC heat capacity and represents its capability to absorb heat 
released from fermion-hole heating.   Note the sensitive dependence 
on mass-ratio illustrated by Fig. 1, where the temperature of mixtures 
that start out at $T=0.01 T_{F}$ is shown for four possible Fermi-Bose 
mixtures with stable alkalis ($^{6}$Li or $^{40}$K for the fermionic 
species, $^{23}$Na or $^{87}$Rb for the bosonic species).
For the sake of comparison, we recover the fermion-hole heating rate 
for a pure fermion system obtained in Ref. \cite{Eddy2} by setting 
$m_{B} \rightarrow 0$, corresponding to a BEC-refrigerator of 
vanishing heat capacity (shown by the thin full line).

\noindent
{\it Trapped interacting mixtures.} -
In addition to the mass-ratio dependence, the spatial density profiles
in the realistic case of trapped gases affect the temperature dynamics.
The fermion-boson overlap can be reduced by a repulsive inter-species
interaction or by the occurrence of a true phase separation.  
In the non-phase separated systems, the mutual overlap can be controlled by
varying the ratio of the fermion to boson trapping frequencies
(using, for instance, a bichromatic trap \cite{Onofrio}).  
A decrease in overlap reduces fermion loss but also reduces the efficiency
of sympathetic cooling.  Which effect wins out, depends sensitively
on the density profiles and, hence, on the interaction parameters. 

\begin{table}
\begin{ruledtabular}

\begin{tabular}{cccccc}
    & s & $^6$Li-$^{87}$Rb & $^6$Li-$^{85}$Rb & $^7$Li-$^{87}$Rb &
$^7$Li-$^{85}$Rb \\
\hline
  $a_T$ & 1.0 & $-43.6$ & $-36.6$ &  $-151$ &  $-118$ \\
    & 0.99280 & $-17.0$ & $-12.9$ & $-60.0$ & $-48.7$ \\
    & 0.97404 & $+17.0$ & $+18.8$ & $+5.43$ & $+8.43$ \\
  $a_S$ & 1.0 &  $+153$ &  $+215$ & $-16.3$ & $-1.11$ \\
    & 0.99280 &  $-152$ & $-77.4$ & $+27.1$ & $+33.8$ \\
    & 0.97404 & $+45.5$ & $+50.4$ &  $+167$ &  $+280$
\end{tabular}
\end{ruledtabular}
\caption{Singlet and triplet elastic scattering lenghts (shifted and 
unshifted) for the mixtures of Li and Rb isotopes, in Bohr radii. 
The values of $s$ are found so that $a_T=\pm 17$ $a_0$, and the same 
scaling was assumed for the singlet. The scattering lengths were 
obtained from potential curves with {\it ab initio} data from \cite{Koek},
joined smoothly to an exponential wall of the form $c e^{-b R}$
at short separations $R$, and to the long-range form 
$-C_{6}/R^{6}-C_{8}/R^{8}-C_{10}/R^{10} \mp AR^\alpha e^{-\beta R}$
at $R$=13.5 $a_{0}$. Here, $\mp$ stands for the singlet $X^{1}\Sigma^{+}$
and triplet $a^{3}\Sigma^+$ molecular states, respectively.
The values for $C_{6}=2545$ a.u. from \cite{Derevianko}, and
$C_{8}=2.34 \times 10^{5}$ a.u. and $C_{10}=2.61 \times 10^{7}$ a.u.
from \cite{Porsev}, were used. The parameters of the exchange energy 
are $\alpha=4.9417$ a.u. and $\beta=1.1836$ a.u. \cite{Smirnov}, while 
the constant $A=0.0058$ a.u. was found by fitting the {\it ab initio} data.}
\label{tab1}
\end{table}

To determine the trap densities in the case of a $^6$Li-$^{87}$Rb
mixture, the most promising candidate shown in Fig. 1, we have 
evaluated the interspecies elastic scattering lengths for 
various Li and Rb isotopes. 
Although spectroscopic data for LiRb potential curves are lacking, 
the excellent agreement between {\it ab initio} \cite{Roos} and 
our parametrized potentials (see Table \ref{tab1}) at large 
separation attest to the accuracy of the curves in that region. 
However, there is a discrepancy between both sets of curves near 
the equilibrium distance. Compared to \cite{Roos} our potential 
curves are deeper by about $3\%$ for the singlet and $7\%$ for 
the triplet channel. By scaling the whole potential curves so 
that $V_s(R)=sV(R)$ with $s\leq 1$, we can explore the effect 
of having shallower potential curves; the values of the scattering 
lengths change significantly as $s$ varies accordingly. 
In the singlet case, one bound level disappears, and the
scattering length varies between $\pm \infty$, while the triplet
scattering length becomes positive. Recently, Zimmermann and co-workers
\cite{Silber} determined the magnitude of the triplet scattering 
length of $^6$Li-$^{87}$Rb to be $|a_T|=17^{+9}_{-6}$ $a_0$.
In Table \ref{tab1}, we present values for the singlet and triplet 
scattering lengths for the unshifted $(s=1)$ and shifted potentials
(so that $a_T$ agrees with $\pm 17$ $a_0$). These values as well as
those calculated in \cite{Ouerdane} are tentative and future measurements 
are required to specify the potentials more accurately.
Such experimental feedback can be obtained from either one of the isotopes 
so that we list the results for all combinations, regardless of their 
fermionic or bosonic nature.

We now study the temperature of a $^{6}$Li -- $^{87}$Rb mixture trapped in an
idealized trap: both species are contained by cylindrical trapping
potentials $V_{{\rm F}}({\bf r})$ and $V_{{\rm B}}({\bf r})$ that are 
of the hard wall type in the transverse direction with the same radius $R_{{\rm T}}$,
where $R_{{\rm T}}$ is much larger than the BEC healing length and the average
fermion-fermion distance. In the longitudinal direction, the fermions
and bosons experience harmonic trapping potentials with different
trapping frequencies $\omega_{{\rm f}}$ and $\omega_{{\rm b}}$ respectively.
Apart from its simplicity, such a situation reproduces the relevant
heating features of a generic three-dimensional cold atom mixture
with anisotropic confinement.  From Table I, we find that some hyperfine
states  of the $^{6}$Li and $^{87}$ Rb isotopes that can be trapped 
magnetically give positive valued inter-species scattering lengths, 
thereby favoring spatial separation in the trap.
The density profiles $n_{{\rm B}}({\bf r})$ and $n_{{\rm F}}({\bf r})$ 
in a Thomas-Fermi approximation are determined by (see Molmer
reference in \cite{lump})
\begin{eqnarray}
n_{{\rm B}}({\bf r}) & = &  \frac{\mu_{{\rm B}}-V_{{\rm B}}({\bf 
r})-\lambda_{{\rm FB}} n_{{\rm F}}
({\bf r})}{\lambda_{{\rm B}}}
\nonumber \\
n_{{\rm F}}({\bf r}) & = & \frac{\left[ 2 m_{{\rm F}} 
\right]^{3/2}}{6 \pi^2 \hbar^3}
\{\mu_{{\rm F}}-V_{{\rm F}}({\bf r}) -
   \lambda_{{\rm FB}} n_{{\rm B}}({\bf r})] \}^{3/2} \ \
\label{densities}
\end{eqnarray}
where $\lambda_{{\rm B}}$ and $\lambda_{{\rm FB}}$ denote the
boson-boson and fermion-boson interaction strengths proportional
to the corresponding $s$-wave scattering lengths,
$a_{B}$ and $a_{FB}$: $\lambda_{{\rm B}}
= 4 \pi \hbar^{2} a_{{\rm B}}/m_{{\rm B}}$ and $\lambda_{{\rm FB}} = 
2 \pi \hbar^{2}
a_{{\rm FB}}\left(1/m_{{\rm F}}+1/m_{{\rm B}} \right)$.  
In  Eq. (\ref{densities}), we tacitly assume temperature independent 
density profiles, a good approximation in the regime of interest. 
Solving Eq. (\ref{densities}) numerically, we find that a decrease 
in confinement strength for the bosons generally increases 
the overlap with the Fermi species.  Controlling the relative trapping 
frequencies allows for a variation of the fermion-boson overlap and, 
hence, the rate of Fermi-hole heating and the efficiency of
sympathetic cooling. The effect of inter-species
interactions is marginal on the Fermi gas but is very pronounced
for the BEC, and the mutual overlap is very sensitive to the
interaction parameters. The temperature trajectories shown
in Fig. 2 are calculated for $\omega_{{\rm f}} = 10^{3} s^{-1}$,
while varying the Bose trap frequency as shown in the caption.
We choose the chemical potentials in the initial state to be 1 $\mu K$
for the Fermi gas and 100 nK for the Bose gas.
The scattering lengths are chosen as $a_{{\rm B}}$=5.8 nm for Rb and 
$a_{{\rm FB}}$=-0.90 nm for the Li-Rb interactions.

We consider the fermions to be subject to loss by background scattering,
$\dot{n}_{{\rm F}}(x) = - \gamma n_{{\rm F}}(x)$, as well as by
the 3-body recombination of Eq. (\ref{3b}).
While these loss-rates are unknown, we have chosen values
that corresponded to  $\gamma=10$ Hz and $\tau^{-1}_{3}
= \alpha n_{B}^{2}=4 s^{-1}$ at the peak densities of case
$\omega_{{\rm f}}/\omega_{{\rm b}}=1$ of Fig. 2. 
The behavior shown in Fig. 2 is quite robust with respect to 
various choices of the relevant parameters.
In addition, we assumed that a continued evaporative cooling
of the bosons drains away energy from the system at a realistic rate.
In the calculations with different fermion and boson trapping frequencies,
we took the evaporative cooling rates to be equal and accounted for the
low overlap drop in evaporative cooling efficiency by introducing a
figure of merit as described in Ref. \cite{Brown}. This allows to 
introduce an effective evaporative cooling rate $-\dot{Q}_{{\rm eff}}$ 
proportional to the overlap of the two clouds.
The rate of free energy change is then identified as sum of three terms:
$dF/dt=\left( dF/dt \right)_{{\rm 3b}} + \left( dF/dt \right)_{{\rm Bkgnd}}
- \dot{Q}_{{\rm eff}}$, where the free energy rates have to be calculated
from the resulting density profiles,
\begin{equation}
\left(\frac{dF}{dt}\right)_{{\rm 3b}}=\frac{(6\pi^2)^{2/3}\hbar^2
    \alpha}
{5m_\mathrm{F}}
\int d^{3}r \; \left[ n_{\mathrm{F}}({\bf r}) \right]^{5/3} \; 
\left[ n_{\mathrm{B}}({\bf r}) \right]^{2} \; ,
\label{3bl}
\end{equation}
whereas as in \cite{Eddy2}, the background scattering free energy
increase rate is given by
\begin{equation}
\left(\frac{dF}{dt}\right)_{{\rm Bkgnd}}=\frac{3 (6\pi^2)^{2/3}
\hbar^{2} \gamma}{10 m_{\mathrm{F}}}
\int d^{3}r \; \left[ n_{\mathrm{F}}({\bf r}) \right]^{5/3} \; .
\label{Bkgndl}
\end{equation}
Finally, the temperature trajectory follows from Eq. (\ref{tempdyn}), in which
we use Eqs. (\ref{3bl}) and (\ref{Bkgndl})
\begin{equation}
\frac{dT}{dt}=\frac{(dF/dt)_{{\rm 3b}}+(dF/dt)_{{\rm 
Bkgnd}}-\dot{Q}_{{\rm eff}}}
{C_{{\rm F}}(T)+C_{{\rm B}}(T)} \; ,
\end{equation}
where $C_{{\rm F}}$ and $C_{{\rm B}}$ denote the fermion and boson 
heat capacities. The resulting temperature dynamics is depicted in 
Fig. 2 for three different trapping frequency ratios. These results 
show that the heating rate is mitigated best by using a larger
$\omega_{\mathrm{f}}/\omega_{\mathrm{b}}$-ratio, at least until the 
spatial overlap is decreased due to an excessive spreading of the Bose cloud.

\begin{figure}[t]
\psfrag{x}[t][]{$t (\mathrm{ms})$}
\psfrag{y}[b][]{$T/T_\mathrm{F}$}
\psfrag{x1}[][]{$\omega_{\mathrm f}/\omega_{\mathrm b}$=1}
\psfrag{x2}[][]{Fermi gas}
\psfrag{x3}[][]{$\omega_{\mathrm f}/\omega_{\mathrm b}$=2}
\psfrag{x4}[][]{$\omega_{\mathrm f}/\omega_{\mathrm b}$=10}
\includegraphics[width=0.90\columnwidth,clip]{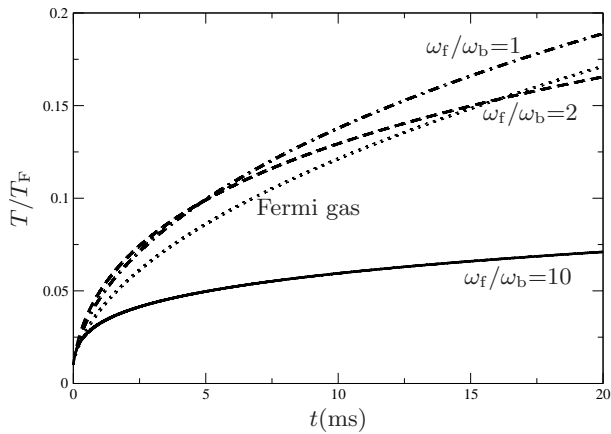}
\caption{Fermi hole heating of harmonically trapped $^6$Li-$^{87}$Rb
mixtures. The dependence of the degeneracy parameter $T/T_\mathrm{F}$
(starting from an initial value of $T/T_\mathrm{F}=10^{-2}$) on time
is depicted for three cases of $\omega_\mathrm{f}/\omega_\mathrm{b}=1$
(dot-dashed curve), $\omega_\mathrm{f}/\omega_\mathrm{b}=2$ (dashed),
$\omega_\mathrm{f}/\omega_\mathrm{b}=10$ (continuous).
The dotted curve represents the intrinsic Fermi hole heating for the same
loss-rate of the fermions but in the absence of the $^{87}$Rb atoms.}
\label{fig2}
\end{figure}

\noindent
{\it Conclusion.} - Our study of loss-induced heating, based on 
an effective equilibration
model, reveals that a cold atom Fermi-Bose mixture of high boson to
fermion mass ratio such as $^6$Li-$^{87}$Rb can be maintained in the 
Fermi liquid temperature regime.
This result, a consequence of the increased BEC heat capacity due 
to the larger trapping frequency ratio between fermions and bosons, 
could open up a new avenue for cold atom studies.
We also discuss a specific cooling strategy which could be soon 
implemented in the experimental investigations going on for the 
specific $^6$Li-$^{87}$Rb mixture, for which anomalous heating
has been observed \cite{Silber}.  

\begin{acknowledgments}
R.C. acknowledges partial support from NSF, R.O. from
Cofinanziamento MIUR, and E.T. from the Los Alamos
Laboratory Directed Research and Development (LDRD) program.
\end{acknowledgments}

\end{document}